\documentclass[prb,twocolumn,showpacs,nobibnotes,nofootinbib,notitlepage]{revtex4}
\usepackage{amsmath}
\usepackage{amssymb}
\usepackage{bm}
\usepackage{epsfig}
\usepackage{graphicx}
\usepackage[english]{babel}
\usepackage[colorlinks=true]{hyperref}

\newcommand{\Tr}{\mathop{\rm Tr}}

\renewcommand{\Im}{\mathop{\rm Im}}

\begin{document}

\newcount\timehh  \newcount\timemm
\timehh=\time \divide\timehh by 60
\timemm=\time
\count255=\timehh\multiply\count255 by -60 \advance\timemm by \count255

\title{Exciton spin noise in quantum wells}

\author{D.S. Smirnov, M.M. Glazov}

\affiliation{Ioffe Physical-Technical Institute of the RAS, 194021, St.-Petersburg, Russia}

%\date{\today, file = \jobname.tex, printing time = \number\timehh\,:\,\ifnum\timemm<10 0\fi \number\timemm}

\begin{abstract}
A theory of spin fluctuations of excitons in quantum wells in the presence of non-resonant excitation has been developed. Both bright and dark excitonic states have been taken into account. The effect of a magnetic field applied in a quantum well plane has been analyzed in detail. We demonstrate that in relatively small fields the spin noise spectrum consists of a single peak centered at a zero frequency while an increase of magnetic field results in the formation of the second peak in the spectrum owing to an interplay of the Larmor effect of the magnetic field and the exchange interaction between electrons and holes forming excitons. 
Experimental possibilities to observe the exciton spin noise are discussed, particularly, by means of ultrafast spin noise spectroscopy.
We show that the fluctuation spectra contain, in addition to individual contributions of electrons and holes, an information about correlation of their spins.
\end{abstract}

\maketitle

\section{Introduction}

The rapid development of experimental techniques of optical spin detection in semiconductors and semiconductor nanostructures has made it possible to apply spin noise spectroscopy for spin dynamics studies in various semiconductor structures.\cite{Zapasskii:13,Oestreich-review,Mueller2010,1742-6596-324-1-012002} This technique was suggested in 1980s to study magnetic resonance in atomic gases by all-optical means, namely by measuring fluctuation spectra of Faraday/Kerr rotation of weak non-resonant probe beam. By now it has already been successfully applied to bulk semiconductors,\cite{PhysRevB.79.035208,Romer2010} multiple and single quantum well structures,\cite{muller-Wells,noise-trions} individual quantum dots and quantum dot ensembles\cite{crooker2010,crooker2012,dahbashi:031906} making it possible to reveal and analyze spin dynamics of electrons and holes.

Nowadays a special interest to the spin fluctuations in non-equilibrium conditions has been formed. On one hand, recent theoretical works have shown that the departure from equilibrium induced by external electric fields\cite{Sinitsyn} or optical pumping\cite{Marina-noise} (see also Ref.~\onlinecite{ivchenko73fluct_eng}) drastically modifies the spin noise spectra. On the other hand, it has been demonstrated experimentally that a probe beam propagating in the region of weak absorption can substantially perturb the spin system.\cite{dahbashi:031906} The effect is most pronounced if special techniques, i.e. use of high-extinction polarization geometries and/or placing the sample into the optical cavity are applied.\cite{Glasenapp,noise-trions} In this regard, the spin noise spectra of excitons can be especially interesting because these particles do not exist without pumping, while their spin dynamics is quite rich, particularly, owing to an interplay of optically active 
and inactive states, see e.g. Refs.~\onlinecite{maialle93,Dareys1993353,Vinattieri93,blackwood94,Vinattiery,marie97_1,toulouse,vina99,PhysRevLett.81.2586,Astakhov07,dyakonov_book,IX_Leonard09,IX_High13}. In Ref.~\onlinecite{Marina-noise} the effects of pumping and Bose-statistics have been studied in detail for exciton-polaritons in microcavity disregarding dark states. Here we study spin noise spectra of non-resonantly created excitons in quantum wells in presence of an external magnetic field with the special emphasis on the interplay of bright and dark states caused by their different lifetimes, exchange interaction, and the magnetic field effects.

\section{Model}
\label{sec-model}

\subsection{General formalism}

We consider a QW grown along $z\parallel [001]$ axis where the heavy-hole excitons are injected by non-resonant optical pumping. In the absence of a magnetic field and an electron-hole exchange interaction the excitonic state is four-fold degenerate in $m_z = s_z + j_z$, being the total spin $z$ component.\cite{ivchenko05a} Here $s_z=\pm 1/2$ is the electron spin $z$ component, $j_z = \pm 3/2$ is the heavy-hole spin $z$ component. Two states with $m_z = \pm 2$ are optically inactive (``dark''), while the doublet with $m_z = \pm 1$ is optically active (``bright''). The isotropic part of electron-hole exchange interaction splits dark and bright doublets by $\hbar\delta_0$, on the order of $100$~$\mu$eV for direct excitons in quantum wells\cite{toulouse} and several $\mu$eV for spatially indirect excitons in single or double quantum wells.\cite{Kesteren90} An in-plane magnetic field mixes bright and dark excitonic states. Hereinafter we assume that the in-plane $g$-factor of heavy-hole is zero,\cite{Mar99} 
hence, the 
transverse magnetic field affects only electron spin state. As a result, it couples the states with $m_z=2$ and $m_z=1$ or with $m_z=-2$ and $m_z=-1$. The excitonic Hamiltonian in the basis of states with $m_z=+2,+1,-1,-2$ reads
\begin{equation}
\label{Hamiltonian}
\mathcal H_x = 
\begin{pmatrix}
-\hbar\delta_0 & \hbar\Omega/2 & 0 & 0\\
\hbar\Omega/2 & 0 & 0 & 0\\
0 & 0 & 0& \hbar\Omega/2 \\
0 & 0 & \hbar\Omega/2 & - \hbar \delta_0
\end{pmatrix}.
\end{equation}
Here we assumed that the external field $\bm B$ is applied along $x$ axis, and introduced the electron Larmor frequency $\Omega  = g_e\mu_B |\bm B|/\hbar$, with $g_e$ being the in-plane electron $g$-factor and $\mu_B$ being the Bohr magneton. In Eq.~\eqref{Hamiltonian} the energy is reckoned from bright doublet. In what follows we neglect the anisotropic (``cubic'') splitting of dark states. For freely propagating excitons, as well as for the excitons localized at the quantum well imperfections the long-range exchange interaction induces the splitting between linearly polarized combinations of bright states,\cite{maialle93,goupalov98} below we assume that the effect of such a splitting can be reduced to the effective exciton  spin-flip time.\cite{maialle93,ivchenko05a} Under these assumptions in order to study the exciton spin dynamics and spin noise spectra it is enough to address the dynamics of the exciton quadruplet, which is described by the $4\times 4$ density matrix $\varrho$ with the elements 
$\varrho_{m_zm_z'}$.
Following Ref.~\onlinecite{toulouse1} we introduce the total number of particles, $N = \Tr\{\varrho\}$, occupancies of the dark and bright excitonic states, $N_d=\varrho_{2,2}+\varrho_{-2,-2}$ and $N_b=\varrho_{1,1}+\varrho_{-1,-1}$, respectively, the electron-in-exciton spin pseudovector $\bm S = (S_x, S_y, S_z) = \Tr{\{\hat{\bm \sigma} \varrho/2\}}$ with $\bm \sigma/2$ being electron spin operator, hole $z$ spin component $J_z = \Tr{\{{\hat J}_z \varrho\}}$ with $\hat J_z$ being the hole spin $z$ component operator, as well as vector $\bm Q = (Q_x, Q_y, Q_z) = \Tr{\{\hat{\bm \sigma} \hat{J}_z \varrho/3\}}$ describing electron-hole correlation. We note that the component $Q_z$ can be expressed as
\begin{equation}
\label{Qz}
Q_z = \frac{N_d - N_b}{2},
\end{equation}
while the in-plane components $Q_x$, $Q_y$ are independent parameters.

In our model, the excitons are assumed to be non-resonantly excited via creation of electron-hole pairs by unpolarized radiation. In the course of energy relaxation electrons and holes loose their spin, as a result, the generation rates of both bright and dark states are the same. Note that in the opposite case of resonant excitation of excitons requires separate study, because in this case the spin polarization and its fluctuations are mainly determined by the driving laser field.

The rate equations describing the dynamics of excitons and their spins are given by, see Ref.~\onlinecite{toulouse1} and Appendix~\ref{appendix} for the details of derivation
\begin{subequations}
 \label{dNbd}
 \begin{equation}
 \label{dNb}
  \frac{d N_b}{d t}=\frac{G}{2}-\frac{N_b}{\tau_R}+Q_z\left(\frac{1}{\tau_h}+\frac{1}{\tau_e}\right)-Q_y\Omega,
 \end{equation} 
 \begin{equation}
  \frac{d N_d}{d t}=\frac{G}{2}-\frac{N_d}{\tau_{NR}}-Q_z\left(\frac{1}{\tau_h}+\frac{1}{\tau_e}\right)+Q_y\Omega,
 \end{equation} 
 \begin{equation}
  \frac{d Q_y}{d t}=-\frac{Q_y}{2\tau_b}-\frac{Q_y}{2\tau_d}-\frac{Q_y}{\tau_e}-\frac{Q_y}{\tau_h}-Q_z\Omega-S_x\delta_0,
 \end{equation} 
 \begin{equation}
  \frac{d S_x}{d t}=-\frac{S_x}{\tau_e}-\frac{S_x}{2\tau_b}-\frac{S_x}{2\tau_d}+Q_y\delta_0.
 \end{equation}
\end{subequations}
Here $G$ is the total generation rate of excitons, $\tau_e$ and $\tau_h$ are the electron and hole spin-flip times, respectively; 
$\tau_b^{-1}=\tau_R^{-1}+\tau_1^{-1}$ and $\tau_d^{-1}=\tau_{NR}^{-1}+\tau_2^{-1}$,  where $\tau_R$ and $\tau_{NR}$ are the times of radiative recombination of bright excitons and non-radiative recombination of dark ones, respectively, and $\tau_1$, $\tau_2$
are the spin-flip times of the bright and dark excitons, respectively. Note, that in the case of nonresonant excitation $\tau_R$ is an effective radiative recombination time which takes into account the process of exciton scattering into the radiative cone.\cite{Vinattiery} Equations~ \eqref{Qz} and \eqref{dNbd} form a closed set providing the mean values of $N_b$, $N_d$, $Q_y$, $Q_z$ and $S_x$ in the presence of unpolarized generation. The remaining quantities $Q_x$, $S_y$, $S_z$ and $J_z$ obey the following equations:
\begin{subequations}
\label{kinetic}
\begin{equation}
\label{Sz}
\frac{ d S_z}{ dt} = \Omega S_y - \frac{S_z}{\tau_e} - \frac{S_z - J_z/3}{2\tau_b} - \frac{S_z + J_z/3}{2\tau_d},
\end{equation}
\begin{equation}
\label{Jz}
\frac{d J_z}{dt} = - \frac{J_z}{\tau_h} - \frac{J_z - 3S_z}{2\tau_b} - \frac{J_z + 3S_z}{2\tau_d},
\end{equation}
\begin{equation}
\label{Sy}
\frac{ d S_y}{ dt} = -\Omega S_z - \delta_0 Q_x- \frac{S_y}{\tau_e} - \frac{S_y}{2\tau_b} - \frac{S_y}{2\tau_d},
\end{equation}
\begin{equation}
\label{Qx}
\frac{ d Q_x}{ dt} = \delta_0 S_y- \frac{Q_x}{\tau_e} - \frac{Q_x}{\tau_h}- \frac{Q_x}{2\tau_b} - \frac{Q_x}{2\tau_d}.
\end{equation}
\end{subequations}
For completeness of the theory we need, along with the the above variables, four more fluctuating quantities, namely $Q_b=\Tr\{(\hat{J}_y\hat{\sigma}_x-\hat{J}_x\hat{\sigma}_y)\varrho/2\}$,  $Q_d=\Tr\{(\hat{J}_y\hat{\sigma}_x+\hat{J}_x\hat{\sigma}_y)\varrho/2\}$, $Q_0=\Tr\{\hat{J}_x\hat{\sigma}_z\varrho/2\}$ and the hole spin component $J_y{=\Tr\{\hat J_y \varrho\}}$. Note that the matrices $\hat J_x$, $\hat J_y$, and $\hat J_z$ differ by the factor $3/2$ from $2\times 2$ Pauli matrices acting in the basis of $j_z=\pm3/2$ hole spin states. The corresponding coupled set of equations reads
\begin{subequations}
\label{kineticQ}
\begin{equation}
 \frac{dQ_b}{dt}=\Omega Q_0-\frac{Q_b}{\tau_b}-\frac{Q_b}{\tau_e}-\frac{Q_b}{\tau_h},
\end{equation} 
\begin{equation}
\frac{dQ_d}{dt}=-\Omega Q_0-\frac{Q_d}{\tau_d}-\frac{Q_d}{\tau_e}-\frac{Q_d}{\tau_h},
\end{equation} 
\begin{equation}
\frac{dQ_0}{dt}=\frac{\Omega}{2}\left(Q_d-Q_b\right)+\frac{\delta_0}{2}J_y-\frac{Q_0}{2\tau_b}-\frac{Q_0}{2\tau_d}-\frac{Q_0}{\tau_e}-\frac{Q_0}{\tau_h},
\end{equation} 
\begin{equation}
\frac{dJ_y}{dt}=-2\delta_0 Q_0-\frac{J_y}{2\tau_b}-\frac{J_y}{2\tau_d}-\frac{J_y}{\tau_h}.
\end{equation} 
\end{subequations}
The sets of equations~\eqref{kinetic} and~\eqref{kineticQ} are decoupled from Eqs.~\eqref{dNbd}. Moreover, Eqs.~\eqref{kineticQ} were not needed in Ref.~\onlinecite{toulouse1} to describe time-resolved polarized photoluminescence, but, as shown below in Sec.~\ref{faraday:ellipticity} can be important in the spin-noise studies. It is noteworthy, that the parameters $Q_x$, $S_y$, $S_z$, $J_z$, $Q_b$, $Q_d$, $Q_0$ and $J_y$ are zero on average, but experience temporal fluctuations.

Our goal is to find spin fluctuation spectra of excitons and, ultimately, to calculate the noise spectra of Faraday, Kerr and ellipticity effects detected by weak probe beam. To describe these fluctuations we introduce a four-component columns $\mathcal Y = (\delta S_z,\delta J_z,\delta S_y, \delta Q_x)$ and $\mathcal Z = (\delta Q_b,\delta Q_d,\delta Q_0, \delta J_y)$ composed of the momentary values of fluctuations and the corresponding correlation functions, for example, $\langle\lbrace\mathcal Y_\alpha(t), \mathcal Y_\beta(t')\rbrace_s\rangle$, where $\alpha,\beta=1,2,3,4$ enumerate the components of vector $\mathcal{Y}$, the curly brackets stand for the symmetrized product $\lbrace\mathcal Y_\alpha(t), \mathcal Y_\beta(t')\rbrace_s=(\mathcal Y_\alpha(t) \mathcal Y_\beta(t') + \mathcal Y_\beta(t') \mathcal Y_\alpha(t)\mathcal)/2$, and the angular brackets denote quantum and ensemble averaging. Under steady state conditions the correlators 
depend on the time difference
$\tau=t-t'$. The fluctuation spectra are given by the Fourier transform of time-dependent correlation functions as\cite{ll5_eng,ll10_eng}
\begin{equation}
 (\mathcal{Y}_\alpha\mathcal{Y}_\beta)_\omega=\int_{-\infty}^\infty\left\langle\left\lbrace\mathcal Y_\alpha(t'+\tau),\mathcal Y_\beta(t')\right\rbrace_s\right\rangle e^{i\omega\tau} d\tau.
\label{spectrum}
 \end{equation} 
We follow general approach to calculate the fluctuations in non-equilibrium conditions\cite{springerlink:10.1007/BF02724353,ll10_eng} and we first solve Eqs.~\eqref{Qz}, \eqref{dNbd} in order to find non-equilibrium steady state occupations of the bright and dark exciton states, $\bar N_b$ and $\bar N_d$, as well as the pseudospin components $\bar S_x$, $\bar Q_y$, $\bar Q_z$. These quantities determine the steady state density matrix of the system, which allows us to find the matrix $\mathcal S\equiv \mathcal S_{\alpha\beta}$ of single-time correlation functions of the fluctuations, $\mathcal S_{\alpha\beta} =\left\langle\mathcal{Y}_\alpha(t)\mathcal{Y}_\beta(t)\right\rangle$. It is given by (the order of fluctuating quantities is, as above, $\delta S_z,\delta J_z,\delta S_y, \delta Q_x$):
\begin{equation}
\label{init}
 \mathcal S=
 \begin{pmatrix}
\bar N/4 & 3 \bar Q_z/2 & - i \bar S_x/2 &  i\bar Q_y/2\\
3 \bar Q_z/2 & 9\bar N/4 & 3\bar Q_y/2 & 3\bar S_x/2\\
 i \bar S_x/2 & 3\bar Q_y/2 & \bar N/4 & - i\bar Q_z/2 \\
- i\bar Q_y/2 & 3\bar S_x/2 &  i\bar Q_z/2 & \bar N/4
\end{pmatrix},
\end{equation} 
with $\bar N = \bar N_b + \bar N_d$. The time-dependent correlation functions $\left\langle\mathcal{Y}_\alpha(t)\mathcal{Y}_\beta(t')\right\rangle$ obey, as functions of $t$ or $t'$, the same set of kinetic equations~\eqref{kinetic} as the fluctuating quantities $\mathcal Y_\alpha(t)$ or $\mathcal Y_\beta(t')$.\cite{Lax1968,ll10_eng,Carmichael} Hence, we represent Eqs.~\eqref{kinetic} in a matrix form, $\dot{\mathcal Y} + \mathcal R\mathcal Y=0$, where dot denotes time derivative and $\mathcal R$ is the matrix describing the a right-hand side of Eqs.~\eqref{kinetic} taken with the opposite sign. It is convenient to calculate directly the spin noise spectra making the unilateral Fourier transform of Eqs.~\eqref{kinetic} with initial conditions given by Eq.~\eqref{init} as
\begin{equation}
\label{chi:s}
(- i \omega + \mathcal R) \chi(\omega) =\mathcal S,
\end{equation}
where the $4\times 4$ matrix $\chi(\omega)$ has the elements ${\chi_{\alpha\beta}(\omega)} =  \int_{0}^\infty\left\langle \mathcal Y_\alpha(t+\tau) \mathcal Y_\beta (t)\right\rangle e^{i\omega\tau} d\tau$. It follows from Eq.~\eqref{chi:s} that $\chi(\omega)=(\mathcal R-  i\omega)^{-1}\mathcal{S}$. Taking into account the symmetry properties of the correlators, the fluctuation spectra can be finally presented as~\cite{ll5_eng,ll10_eng}
\begin{equation}
\label{corr_def}
 (\mathcal{Y}_\alpha\mathcal{Y}_\beta)_\omega= {\frac{1}{2}}\left[\chi_{\alpha\beta}(\omega)+\chi^*_{\beta\alpha}(\omega)+\chi^*_{\alpha\beta}(-\omega)+\chi_{\beta\alpha}(-\omega)\right],
\end{equation} 
where  asterisk denotes the complex conjugate. Similar procedure is used to calculate the fluctuations spectra of $Q_b$, $Q_d$, $Q_0$ and $J_y$, $(\mathcal Z_\alpha Z_\beta)_\omega$ as well as cross-correlations, e.g., $(\mathcal Y_\alpha \mathcal Z_\beta)_\omega$. Note, that as follows from Eq.~\eqref{corr_def} 
the frequency spectra $(\mathcal Y_\alpha^2)_\omega$ are real and $(\mathcal Y_\alpha \mathcal Y_\beta)_\omega=(\mathcal Y_\beta \mathcal Y_\alpha)_\omega^*$, while the observable spectra contain real combinations, $(\mathcal Y_\alpha \mathcal Y_\beta)_\omega+(\mathcal Y_\beta \mathcal Y_\alpha)_\omega$, see Sec.~\ref{faraday:ellipticity}.

\subsection{Fluctuations of spin-Faraday and ellipticity effects}\label{faraday:ellipticity}

In the spin noise spectroscopy experiments the spin Faraday, Kerr or ellipticity signals are measured by weak linearly polarized probe beam propagating along the growth axis $z$.\cite{Zapasskii:13,Oestreich-review,Mueller2010,1742-6596-324-1-012002} {For narrow enough quantum well one can represent the transmitted, $\bm E_{t}$, and reflected, $\bm E_{r}$, beams as
\begin{equation}
\label{fields}
\bm E_{t} = \hat t \bm E_0, \quad \bm E_{r} = (\hat t - 1) \bm E_0,
\end{equation}
where $\bm E_0$ is the amplitude of the \emph{cw} incident light, and $\hat t$ is the matrix of transmission coefficients.\cite{ivchenko05a} The time-average values of its matrix elements are in the basis of $\sigma^\pm$ polarized states (neglecting static anisotropy of the quantum well)  $\bar t_\pm = \bar t_\mp =0$, $\bar t_+ = \bar t_- = t$, with\cite{ivchenko05a}
\begin{equation}
\label{transmission}
 t=\frac{\omega_0 - \omega -i\Gamma}{\omega_0-\omega-i(\Gamma+\Gamma_0)},
\end{equation} 
where $\omega_0$ is the exciton resonance frequency (assumed to be close to $\omega$), $\Gamma_0$ is the exciton radiative recombination rate, $\Gamma$ is its nonradiative damping rate.  Due to excitonic fluctuations in the quantum well, the matrix elements of $\hat t$ change with time yielding measurable noise of the transmitted/reflected light. Particularly, fluctuations of spin-$z$ components of electrons and holes give rise to 
\begin{subequations}
\label{t:fluct}
\begin{equation}
\label{tc:fluct}
\delta (t_+ - t_-) \equiv \delta t_c = a\delta S_z + b\delta J_z,
\end{equation}
while fluctuations of excitonic linear polarization $\delta P_l$ and $\delta P_l'$ in the axes $(x,y)$ and $(x',y')$ rotated by $45^\circ$ result in fluctuations of 
\begin{equation}
\label{tl:fluct}
-\delta (t_\pm + t_\mp) \equiv \delta t_l  \propto \delta P_l,
\end{equation}
\begin{equation}
\label{tl1:fluct}
\mathrm i \delta (t_\pm - t_\mp) \equiv \delta t_l' = c \delta Q_b \propto \delta P_l'.
\end{equation}
\end{subequations}
%\todo{Check signs and consistency}
Here $a$, $b$, and $c$ are constants determined by microscopic mechanism of transmission modulation. For instance, if excitonic fluctuations result, due to exciton-exciton interactions, in the fluctuations of exciton resonance energy then in Eq.~\eqref{transmission} $\omega_0$ should be replaced by $2\times2$ matrix $\hat{\omega}_0$ with the elements $\omega_+,\omega_-,\omega_\pm$ and $\omega_\mp$. These elements obey relations $\omega_++\omega_-=2\omega_0$ and $\omega_\pm=\omega_\mp^*$; the constants in Eqs.~\eqref{t:fluct} can be expressed as
%resonance frequency $\omega_0$ these constants are given by
\begin{equation}
\label{abc}
a = 2\frac{\partial \omega_+}{\partial S_z} \frac{\partial t}{\partial \omega_0},~~b = 2\frac{\partial \omega_+}{\partial J_z} \frac{\partial t}{\partial \omega_0},~~c = i\frac{\partial (\omega_\pm-\omega_\mp)}{\partial Q_b} \frac{\partial t}{\partial \omega_0},
\end{equation} 
where $\partial t/\partial\omega_0 =   -i \Gamma_0/[\omega_0 - \omega - i (\Gamma_0 + \Gamma)]^2$. The detailed calculation of the derivatives such as $\partial \omega_{+}/\partial S_z$ is beyond the scope of the present paper, see Refs.~\onlinecite{Zhukov07,glazov:review,shen05,Nalitov14} for details. It is noteworthy that the parameters $a$, $b$ and $c$ contain both real and imaginary parts.}

For the probe beam polarized along $x$-axis the small fluctuations of Faraday and ellipticity angles $\delta\theta_F$ and $\delta\theta_E$ are given by\cite{glazov:review,noise-trions,gi2012noise}
\begin{equation}
 \delta\theta_E+i\delta\theta_F=\frac{\delta t_c-i\delta t_l'}{2t}.
\end{equation} 
Making use of Eqs.~\eqref{abc} the fluctuations can be presented as
\begin{equation}
 \delta\theta_E+i\delta\theta_F=A\delta S_z+B\delta J_z+C\delta Q_b,
 \label{AB}
\end{equation} 
where we have introduced another three complex parameters 
\begin{equation}
\label{ABC:micro}
A=a/(2t), \quad B=b/(2t), \quad C=-ic/(2t),
\end{equation}
 describing
the sensitivity of Faraday rotation and ellipticity to electron and hole spin polarizations and to exciton linear polarization fluctuations in the $(x',y')$ axis $\delta Q_b\propto \delta P_l'$.

It follows from Eqs.~\eqref{AB}, \eqref{ABC:micro} that, in general, both Faraday rotation and ellipticity fluctuations are contributed by the spin fluctuations of carriers and by fluctuations of exciton linear polarization. In the region of weak absorption, where $|\omega - \omega_0| \gg \Gamma_0, \Gamma$, the fluctuations of Faraday rotation are dominated by the fluctuations $\delta S_z$, $\delta J_z$, while ellipticity fluctuations are mainly determined by $\delta Q_b$. For relatively small detuning of the probe beam from excitonic resonance the contributions of $\delta Q_b$ to Faraday rotation fluctuations and of spin fluctuations to ellipticity noise increase. Note, that the Kerr rotation is, generally, a superposition of Faraday and ellipticity signals,\cite{glazov:review} its noise is given by similar to Eq.~\eqref{AB} expression. From the symmetry point of view contributions $\propto A,B$ are related with circular birefringence/dichroism, while the contributions $\propto C$ are 
related with 
linear birefringence/dichroism.

The fluctuation spectrum of Faraday rotation, $(\delta\theta_F^2)_\omega$, is defined similarly to spin noise spectrum Eq.~\eqref{spectrum}
\begin{equation}
\label{spectrumF}
 (\delta\theta_F^2)_\omega=\int_{-\infty}^\infty\left\langle\{{\delta}\theta_F(t+\tau){,\delta}\theta_F(t)\}_s\right\rangle e^{i\omega\tau} d\tau.
\end{equation} 
For the sake of simplicity in what follows we consider separately contributions of spin fluctuations and excitonic linear polarization fluctuations. Particularly, for the Faraday rotation in the region of relative transparency $\Im C\equiv 0$ and the noise spectrum consists of three contributions:
\begin{equation}
\label{3contrib}
 (\delta\theta_F^2)_\omega=A_F^2(\delta S_z^2)_\omega+B_F^2(\delta J_z^2)_\omega+A_FB_F\mathcal{C}_\omega,
\end{equation} 
where $A_F=\Im A$ and $B_F=\Im B$. The first two terms are responsible for individual contributions of electron and hole spin noise, $(\delta S_z^2)_\omega=(\mathcal Y_1^2)_\omega$ and $(\delta J_z^2)_\omega=(\mathcal Y_2^2)_\omega$, respectively. Besides these individual contributions, the fluctuations of measured Faraday rotation contain the contribution 
of the cross correlations between electrons and holes $\mathcal{C}_\omega \equiv 2(\left\lbrace\delta S_z, \delta J_z\right\rbrace_s)_\omega = (\mathcal Y_1 \mathcal Y_2)_\omega + (\mathcal Y_2 \mathcal Y_1)_\omega$ given by the third term in Eq.~\eqref{3contrib}.

\section{Results and discussion}

In this section we present and analyze the results of spin noise spectra calculations. Table~\ref{tab:params} summarizes the parameters governing exciton spin dynamics taken from Refs.~\onlinecite{maialle93,Vinattieri93,Vinattiery,vina99}. One can see that for the spatially direct excitons the four shortest timescales are determined by the exchange splitting of dark and bright states, hole and exciton spin flips as well as exciton radiative decay.
While $\delta_0^{-1}$ is the smallest parameter, the particular relation between $\tau_h$, $\tau_1$, $\tau_R$ depends on the sample parameters, excitation conditions, etc. We stress that the values presented in Tab.~\ref{tab:params} can be considered as illustrative ones only, because the spin-flip and radiative decay times strongly depend on the exciton distribution, which is determined by the experimental conditions being different for the photoluminescence and spin-noise measurements.

\begin{table}
\caption{The parameters governing direct exciton spin dynamics in GaAs-based quantum wells. Presented data correspond to the mean values of the parameter ranges reported in corresponding references.}
\label{tab:params}
\begin{center}
\begin{ruledtabular}
\small{
\begin{tabular}{c|c|c|c|c|c|c}
 Reference & $\hbar\delta_0$, $\mu$eV & $\delta_0^{-1}$, ps & $\tau_h$, ps & $\tau_{1}$, ps & $\tau_R$, ps & $\tau_e$, ps\\
 \hline
\onlinecite{maialle93} (model) & 100 & 6.6 & 100 & 50 & 400 & 200\\
\hline
\onlinecite{maialle93} (exper.) & 250 & 2.6 & 20 & 65 & 250 & 300\\
\hline
\onlinecite{Vinattieri93}\footnote{The parameters are given for zero electric field.} & -- & -- & 100 & 50 & 300 & 3300\\
\hline
\onlinecite{Vinattiery} & 100 & 6.6 & 125 & 75 & 60 & 1000  \\
\hline
\onlinecite{vina99} & 80 & 8.2 & 25 & 50 & 230\footnote{Extracted from Fig. 3 of Ref.~\onlinecite{vina99}.} & 320
\end{tabular}
}
\end{ruledtabular}
\end{center}
\end{table}

\begin{figure}[tb]
\includegraphics[width=\linewidth]{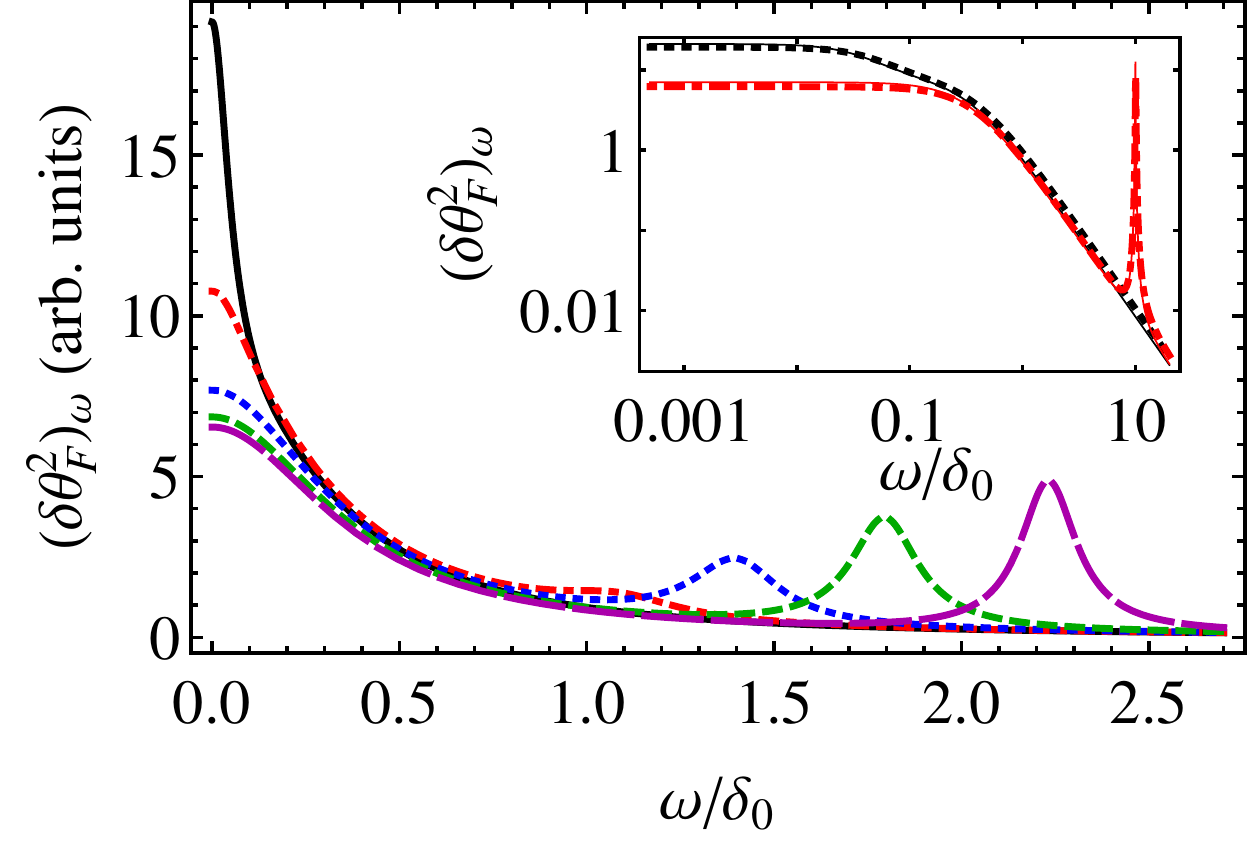}
\caption{(Color online) Spin noise spectrum for different external magnetic fields: $\Omega=0$ (black solid line), $\delta_0/2$ (red dash-dotted line), $\delta_0$ (blue dots), $1.5\delta_0$ (green short-dashed line), $2\delta_0$ (magenta long-dashed line) calculated after Eq.~\eqref{3contrib}. 
The inset shows the spectra at $\Omega=0$ (black dots), $10\delta_0$ (red dash-dotted line) in doubly logarithmic scale. The results of analytical calculation after Eqs.~\eqref{zeroB}, \eqref{highB} are shown by thin curves. The parameters are as follows $\delta_0=0.3$~ps$^{-1}$, $\tau_h=10$~ps, $\tau_R=70$~ps, $\tau_1=100$~ps, $\tau_e=200$~ps, $\tau_2=\tau_{NR}=5$~ns, $B_F = -A_F/2$.}
\label{fig-spectra}
\end{figure}

General equations~\eqref{kinetic}, \eqref{kineticQ}, and \eqref{corr_def}
provide the full description of exciton spin noise spectra in our model. The analytical result is too cumbersome to be presented here. Hence, we demonstrate the numerical results for a generic set of parameters and provide analytical formulae for specific limiting cases. The Faraday-rotation noise spectra calculated numerically after Eqs.~\eqref{corr_def}, \eqref{3contrib} after Eq.~\eqref{spectrumF} for the probe beam propagating in the transparency region are shown in Fig.~\ref{fig-spectra} for different magnetic fields in the case of direct excitons. The parameters of calculation are presented in the Figure caption. In the absence of magnetic field and in moderate magnetic field $\Omega\lesssim\delta_0$ there is one peak in the spin noise spectrum at $\omega=0$. It gets lower and wider with an increase of the field. For higher magnetic fields the second peak appears at the combination frequency 
\begin{equation}
\label{omega:comb}
\omega \approx \Omega' \equiv \sqrt{\Omega^2+\delta_0^2}
\end{equation} 
caused by exciton spin beats.\cite{toulouse,toulouse1} Its position shifts to higher frequencies with an increase of the field while its amplitude increases with the field and then saturates.

\begin{figure}[tb]
\includegraphics[width=\linewidth]{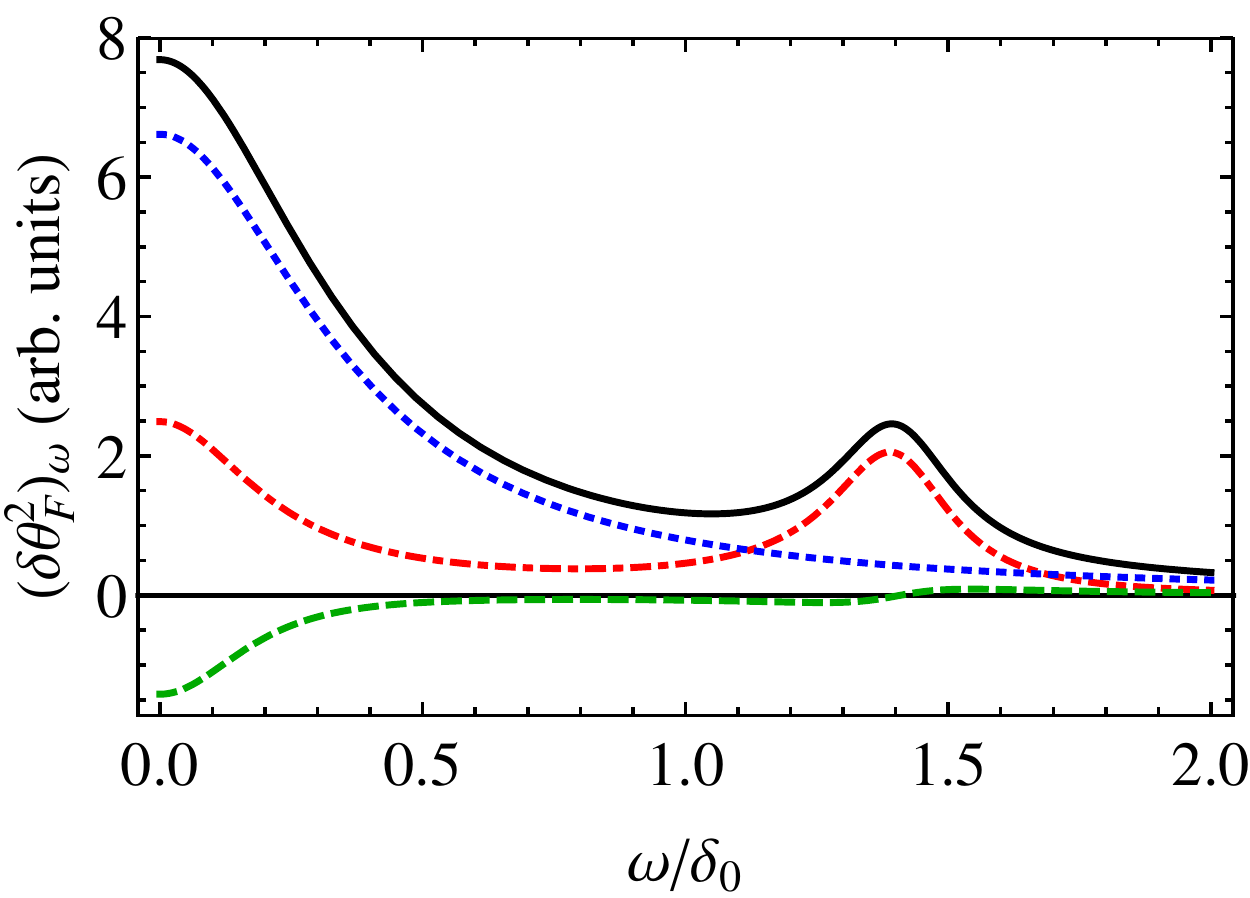}
\caption{(Color online) The spin noise spectrum of excitons (black solid line) and the partial contributions related with electrons (red dash-dotted line), holes (blue dots) and correlation between them (green dashed line). The parameters of calculation are the same as in Fig.~\ref{fig-spectra} and $\Omega=\delta_0$.}
\label{fig-contrib}
\end{figure}

Figure~\ref{fig-contrib} demonstrates three contributions to the Faraday rotation noise spectra caused by: (i) electron spin fluctuations, $(\delta S_z^2)_\omega$, (ii) hole spin fluctuations, $(\delta J_z^2)_\omega$, and (iii) the electron-hole correlation contribution, $\mathcal C_\omega$, calculated numerically for the intermediate value of magnetic field $\Omega=\delta_0$. Interestingly, all the contributions are of importance for quantitative description of zero-frequency peak. In the vicinity of the spin precession peak at $\omega = \Omega'$ the main contribution to the spin noise comes from the electrons. The correlation noise $\mathcal C_\omega$ provides a small and asymmetric contribution, while the hole spin fluctuation spectrum has a small dip at $\omega \approx \Omega'$ (not seen in the used scale). These minor contributions, however, do not significantly modify the overall spin noise spectrum in the vicinity of the spin precession peak.

Now we turn to the detailed analysis of the spin noise spectra. For the direct excitons it is reasonable to assume that
\begin{subequations}
\label{assumptions}
\begin{equation}
\delta_0^{-1}\ll\tau_h,\tau_b,\tau_R\ll\tau_e,\tau_d.
\label{assumptions1}
\end{equation}
If, additionally,
\begin{equation}
\tau_h \ll \tau_b,
\label{assumptions2}
\end{equation}
\end{subequations}
the numbers of bright and dark excitons are equalized by fast hole spin flips, $\bar N_b =\bar N_d = \bar N/2$. Moreover, the hole-in-exciton spin fluctuations become practically independent from the electron-in-exciton spin fluctuations. In the absence of magnetic field one has using Eqs.~\eqref{Sz} and \eqref{Jz}:
% \begin{subequations}
% \label{zeroB}
% \begin{equation}
%   (\delta S_z^2)_\omega=\frac{\bar N\tau_b}{1+4\omega^2\tau_b^2}, \quad  (\delta J_z^2)_\omega=\frac{9}{2}\frac{\bar N\tau_h}{1+\omega^2\tau_h^2}.
% \end{equation}
%  \begin{equation}
%  \label{zeroB:C}
%   \mathcal{C}_\omega=\frac{3\bar N\tau_h}{2\tau_R}\left[\frac{\tau_b+2\tau_R}{1+4\omega^2\tau_b^2} + \frac{\tau_h}{4\tau_b}\frac{\tau_b-2\tau_R}{1+\omega^2\tau_h^2}\right].
%  \end{equation}
%  \end{subequations}
\begin{equation}
\label{zeroB}
  (\delta S_z^2)_\omega=\frac{\bar N\tau_b}{1+4\omega^2\tau_b^2}, \quad  (\delta J_z^2)_\omega=\frac{9}{2}\frac{\bar N\tau_h}{1+\omega^2\tau_h^2}.
\end{equation}
As a result at low frequencies $\omega \lesssim 1/\tau_b$ the spin noise spectrum is dominated by electrons and it has a Lorentzian form with the half-width at the half-maximum $1/(2\tau_b)$. This peaked contribution of the electrons is superimposed at smoother background caused by the hole spin noise. The correlation noise $\mathcal{C}_\omega$ is negligible in the case of fast hole spin flips. The spin noise spectrum calculated after Eqs.~\eqref{zeroB} reproduces well the results of numerical calculation at $\Omega=0$, see black dots and black solid line in the inset in Fig.~\ref{fig-spectra}.

In the presence of sufficiently strong magnetic field, where in addition to Eqs.~\eqref{assumptions}
\[
\Omega^{-1}\ll \delta_0^{-2}\tau_b^{-1} \ll \delta_0^{-1},
\]
the exchange interaction between the electron and hole is not important, correlation fluctuations are absent, $\mathcal C_\omega=0$, while electron and hole contributions to the spin noise spectrum assume the form
\begin{subequations}
\label{highB}
\begin{equation}
\label{highB:e}
  (\delta S_z^2)_\omega=\frac{\bar N}{2}\left[\frac{\tau_b}{1+4(\omega-\Omega)^2\tau_b^2}
  +\frac{\tau_b}{1+4(\omega+\Omega)^2\tau_b^2}
  \right],
\end{equation} 
\begin{equation}
\label{highB:h}
  (\delta J_z^2)_\omega=\frac{9}{2}\frac{\bar N\tau_h}{1+\omega^2\tau_h^2}.
\end{equation}
\end{subequations}
Clearly the electron spin precession gives rise to the Lorentzian-shaped peak in the spin noise spectrum at the Larmor frequency, while the hole spin relaxation contributes to the zero frequency peak. The widths of the peaks are determined by the effective relaxation times $2\tau_b$ and $\tau_h$, respectively. Analytical equations~\eqref{highB} are in a good agreement with numerical results, see red solid line and dash-dotted line the inset in Fig.~\ref{fig-spectra}, which present the spin noise spectrum for $\Omega=10\delta_0$.

Now we turn to the case of slow hole spin relaxation, which can be realized in relatively narrow quantum wells and/or under quasi-resonant excitation. Correspondingly, we assume that Eq.~\eqref{assumptions1} holds, while instead of Eq.~\eqref{assumptions2} we have 
\[
\tau_b\ll\tau_h.
\]
In the absence of magnetic field or in a weak field, $\Omega \ll \tau_R^{-1}$ the excitons can recombine before hole undergoes a spin flip. As a result,\cite{fnote}
\begin{equation}
 \frac{N_d}{N_b}=1+\frac{\tau_h}{\tau_R}.
\end{equation} 
The spin noise spectra have independent contributions of the dark and bright excitons:
\begin{subequations}
 \label{zeroB2}
 \begin{equation}
  (\delta S_z^2)_\omega=\frac{\bar N_d\tau_h}{1+4\omega^2\tau_h^2} + \frac{\bar N_b(2\tau_R-\tau_b)\tau_b}{4\tau_R(1+\omega^2\tau_b^2)}, 
 \end{equation} 
 \begin{equation}
  (\delta J_z^2)_\omega=\frac{9\bar N_d\tau_h}{1+4\omega^2\tau_h^2} + \frac{9\bar N_b(2\tau_R+\tau_b)\tau_b}{4\tau_R(1+\omega^2\tau_b^2)},
 \end{equation} 
 \begin{equation}
  \mathcal{C}_\omega=\frac{6\bar N_d\tau_h}{1+4\omega^2\tau_h^2} - \frac{3\bar N_b\tau_b}{1+\omega^2\tau_b^2}.
 \end{equation}
\end{subequations}
The contribution of bright excitons has a width $1/\tau_b$ while the dark states provide a peak with the width $1/(2\tau_h)$.
We also stress that at $\omega\gtrsim1/\tau_R$ the contributions of the bright and dark states can be of the same order of magnitude.
In high magnetic field, $\Omega \gg \delta_0^2\tau_b \gg \delta_0$, the electron-hole correlation gets suppressed. Consequently, we arrive at Eq.~\eqref{highB:e} for $(\delta S_z^2)_\omega$ and the hole spin fluctuations are described instead of Eq.~\eqref{highB:h} by $(\delta J_z)^2_\omega = 9\bar N\tau_b/(1+4\omega^2\tau_b^2)$.

\begin{figure}[t]
\includegraphics[width=\linewidth]{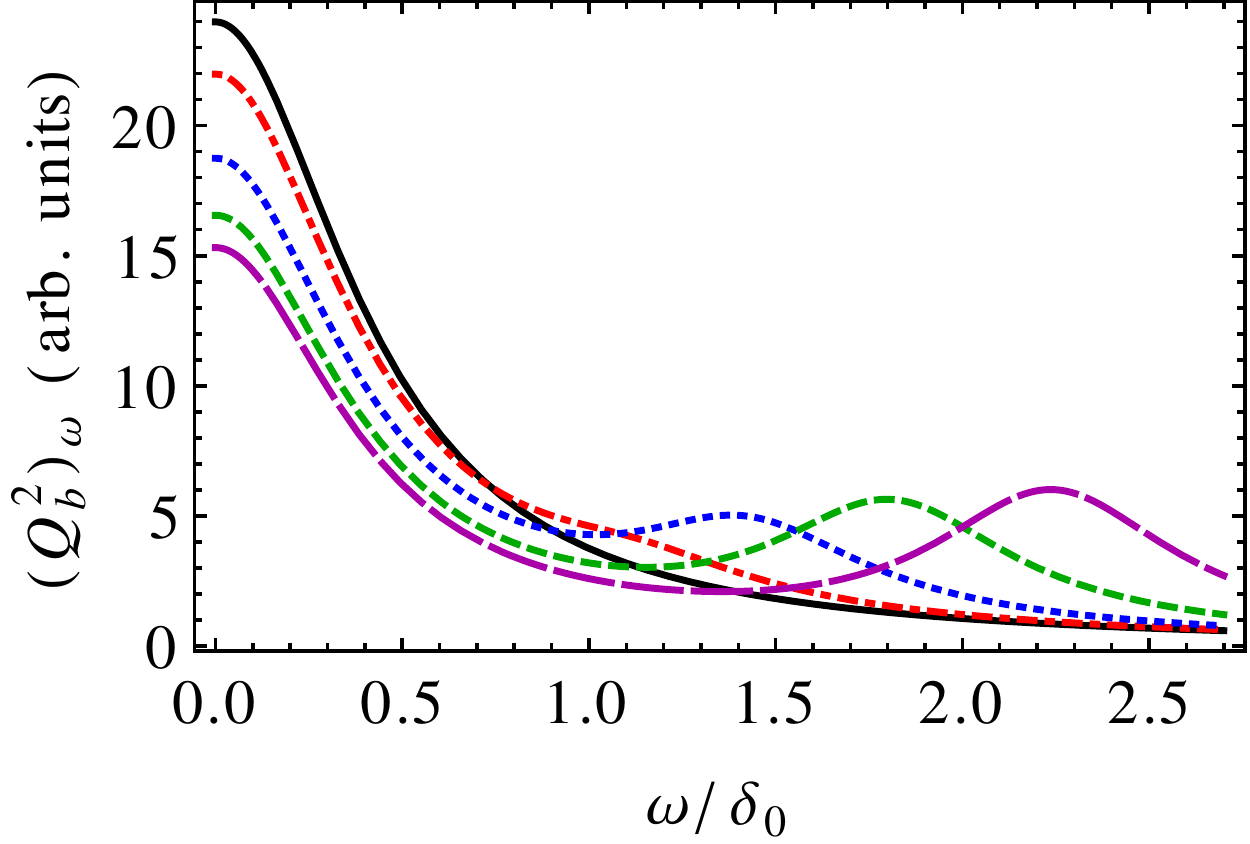}
\caption{(Color online) 
Fluctuation spectrum of exciton linear polarization, $(\delta Q_b^2)_\omega$, for different external magnetic fields: $\Omega=0$ (black solid line), $\delta_0/2$ (red dash-dotted line), $\delta_0$ (blue dots), $1.5\delta_0$ (green short-dashed line), $2\delta_0$ (magenta long-dashed line). The parameters are the same as in Fig.~\ref{fig-spectra}.}
\label{fig-ell}
\end{figure}

Now we briefly analyze the contribution of the exciton linear polarization to the spin signals fluctuations spectra, which manifest themselves, e.g., as fluctuations of ellipticity for the detuned from exciton resonance probe beam. The noise spectrum of $(Q_b^2)_\omega$ in the absence of magnetic field is a Lorentzian centered at zero frequency. It has a width $1/\tau_b+1/\tau_e+1/\tau_h$, roughly corresponding to the shortest spin relaxation times in the system. The transformation of the spectrum with the magnetic field shown in Fig.~\ref{fig-ell} is similar to the electron and hole spin noise [Fig.~\ref{fig-spectra}] with the difference that the width of the peaks are larger in the case of exciton linear polarization fluctuations.

Interestingly, the peak centered at zero frequency does not disappear in high magnetic fields (note that according to Eqs.~\eqref{kineticQ} $Q_b+Q_d$ dose not experience temporal oscillations). Another peak, which in relatively high magnetic fields shifts $\propto \Omega'$, Eq.~\eqref{omega:comb}, is mainly caused by the electron-in-exciton spin precession.

\begin{figure}[t]
\includegraphics[width=\linewidth]{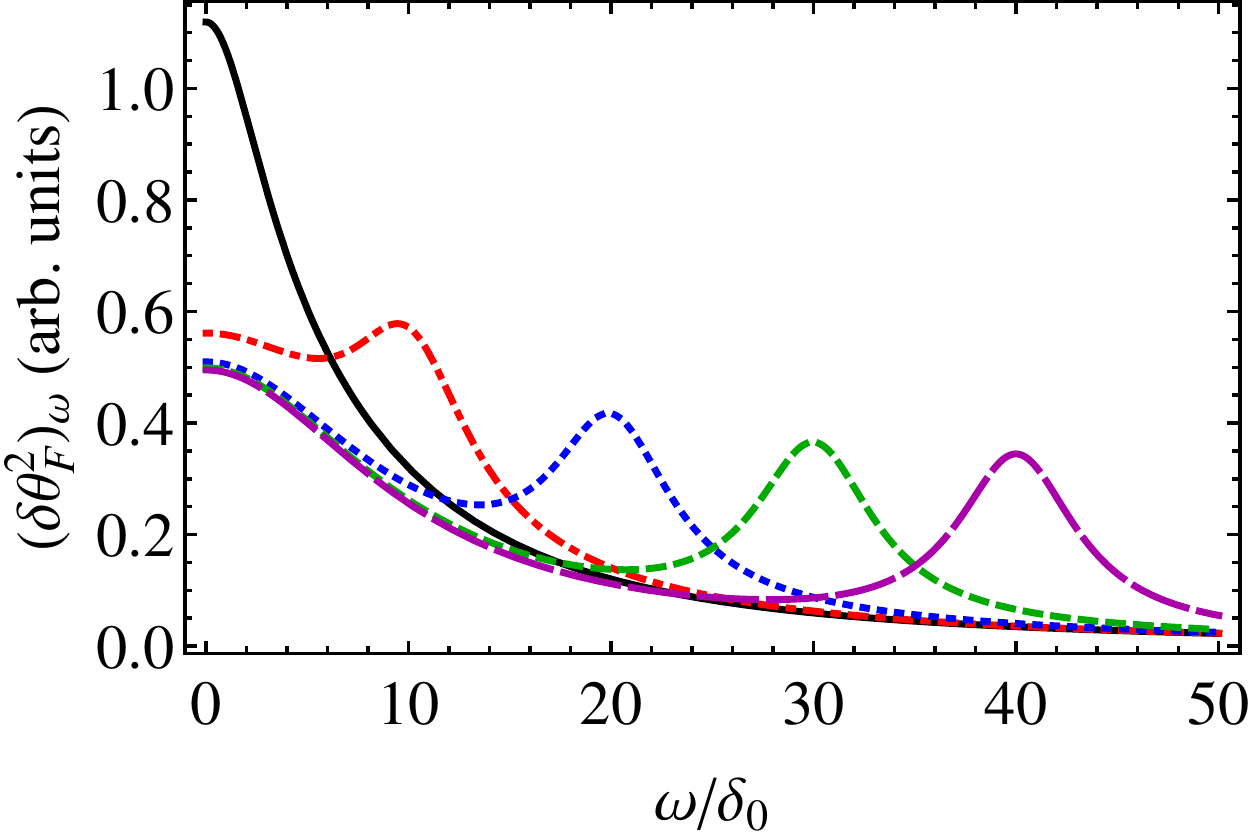}
\caption{(Color online) Spin noise spectrum of indirect excitons for different external magnetic fields: $\Omega=0$ (black solid line), $10\delta_0$ (red dash-dotted line), $20\delta_0$ (blue dots), $30\delta_0$ (green short-dashed line), $40\delta_0$ (magenta long-dashed line). The parameters are as follows $\delta_0=1$~ns$^{-1}$, $\tau_h=0.1$~ns, $\tau_e=0.3$~ns, $\tau_R=\tau_{NR}=\tau_1=\tau_2=10$~ns, $B_F = -A_F/2$.
}
\label{fig-IX}
\end{figure}

Finally we address the case of spatially indirect excitons. Due to small spatial overlap of electron and hole wavefunctions the radiative lifetime is strongly enhanced and dark-bright splitting $\delta_0$ is strongly reduced. Then it is reasonable to assume that the electron and hole spin flips times (tenths on nanoseconds) are the shortest in the system\cite{IX_Leonard09,IX_High13,IX_Kinetics}, namely
\[
\tau_e, \tau_h \ll \delta_0^{-1}, \tau_R, \tau_1, \tau_2, \tau_{NR},
\] 
which allows us to treat the electron and hole spin fluctuations independently. In such a limit again $\bar N_b = \bar N_d = \bar N/2$ and the correlation contribution to the spin noise spectrum is absent. Neglecting dark-bright splitting $\delta_0$ completely we obtain for the whole magnetic field range Eq.~\eqref{highB:h} for the hole contribution to the spin noise spectrum and Eq.~\eqref{highB:e} with the replacement $\tau_b\to \tau_e/2$ for the electron contribution. Hence hole contribution is always centered at $\omega=0$ while the electron spin precession peak follows the electron Larmor frequency.\cite{braun2007,kos2010,gi2012noise} This is illustrated in Fig.~\ref{fig-IX}, where the numerically calculated spin noise spectrum is presented and the transformation of the single peak at zero magnetic field to the two-peak structure is seen. The difference between numerical results and analytical expression does not exceed 5\% for these conditions.

\section{Conclusions}

In the present paper we have developed a theory of spin noise of quantum well excitons for the non-equilibrium conditions. 
The interplay between the exchange splitting of the dark and bright excitonic states and the Zeeman effect of the external magnetic field determines the shape of the spin noise spectrum, which consists (for positive frequencies) of two peaks. One of the peaks centered at $\omega=0$ somewhat decreases with the field, while the peak at the combination frequency $\sqrt{\Omega^2+\delta_0^2}$ is due to the spin precession and appears at relatively strong magnetic fields, $\Omega\gtrsim \delta_0$. For the spatially indirect excitons, the spin noise spectrum also has two peaks, the one at $\omega=0$ is related with the hole spin fluctuations, while the other at $\omega \approx \Omega$ is related with the electron spin noise. The numerical calculations of the spin noise spectra are well described by simple analytical asymptotics.
The spin noise spectra are strongly sensitive to the exchange interaction, magnetic field and relaxation times in the system.
The spin noise technique being sensitive to the spin dynamics of dark and bright excitons already in the absence of magnetic field or in weak magnetic fields as well as to the correlation function of electron and hole spins  may provide complementary information to the conventional time-resolved photoluminescence spectroscopy  or pump-probe technique (see e.g. Refs.~\onlinecite{toulouse,vina99,Hagele,toulouse1}).

The spin fluctuations of indirect excitons can be studied by conventional spin-noise technique owing to relatively long lifetimes and weak exchange interaction, since $\Omega \sim \delta_0 \sim 1$~ns$^{-1}$ (corresponds to $B\sim30$ mT). The power noise spectrum of the direct excitons corresponds to sub-terahertz frequency range. In this case it might be preferable to use advanced methods such as ultrafast spin noise spectroscopy.\cite{digitizing,starosielec:051116,Mueller2010} Particularly, using the sequence of short pulses would enable to measure direct exciton spin noise limited only by the inverse pulse duration, $\sim 1$~ps$^{-1}$.\cite{Berski-fast-SNS}

\acknowledgements
We are grateful to E.L. Ivchenko, A.V. Kavokin and V.S. Zapasskii for valuable discussions. This work was partially supported by Russian Science Foundation (project 14-12-01067), DSS is grateful to Dynasty Foundation.

%\newpage
\appendix
\section{Derivation of kinetic equations}\label{appendix}

The spin dynamics of excitons are described by the $4\times 4$ density matrix $\varrho$ with the elements $\varrho_{m_zm_z'}$. It obeys the master equation
\[
 \dot\varrho(t)=-\frac{i}{\hbar}\left[\mathcal H_x,\varrho(t)\right]-\mathcal L\left\{\varrho(t)\right\},
\]
where the Hamiltonian $\mathcal H_x$ is given by Eq.~\eqref{Hamiltonian}, and the linear operator $\mathcal L\left\{\varrho(t)\right\}$ describes incoherent relaxation precesses and excitons generation. 
%As an example we will derive equations~\eqref{dNb} and~\eqref{Sz}:
%\[
%  \frac{d N_b}{d t}=\frac{G}{2}-\frac{N_b}{\tau_R}+Q_z\left(\frac{1}{\tau_h}+\frac{1}{\tau_e}\right)-Q_y\Omega,
%\]
%\[
% \frac{ d S_z}{ dt} = \Omega S_y - \frac{S_z}{\tau_e} - \frac{S_z - J_z/3}{2\tau_b} - \frac{S_z + J_z/3}{2\tau_d}.
%\]
%The systems~\eqref{dNbd}, \eqref{kinetic} and~\eqref{kineticQ} can be derived in the same way.

%Number of bright excitons is $N_b=\varrho_{+1,+1}+\varrho_{-1,-1}$.

{The precession terms result from the commutator, $[\mathcal H_x,\varrho(t)]$, the nonresonant generation results in the incoming terms $\mathcal L_{+1,+1}^{in} = \mathcal L_{-1,-1}^{in} = \mathcal L_{+2,+2}^{in} = \mathcal L_{-2,-2}^{in} = -G/4$.} Each of elements $\varrho_{+1,+1},~\varrho_{-1,-1}$ decays due to the radiative recombination of the excitons, so the term $-N_b/\tau_R$ should be included in the rate equation. The spin relaxation of bright excitons does not affect their total number. The electron and the hole spin flips invert the spin of one of the carriers and intermix bright and dark excitons. The corresponding spin flips result into terms:
\[
 (\dot\varrho_{+1,+1})_{sf}=-\varrho_{+1,+1}\left(\frac{1}{\tau_e}+\frac{1}{\tau_h}\right)+\frac{\varrho_{+2,+2}}{\tau_e}+\frac{\varrho_{-2,-2}}{\tau_h},
\]
\[
 (\dot\varrho_{-1,-1})_{sf}=-\varrho_{-1,-1}\left(\frac{1}{\tau_e}+\frac{1}{\tau_h}\right)+\frac{\varrho_{-2,-2}}{\tau_e}+\frac{\varrho_{+2,+2}}{\tau_h}.
\]
Adding up these equations and taking into account that $Q_z=(\varrho_{+2,+2}-\varrho_{+1,+1}-\varrho_{-1,-1}+\varrho_{-2,-2})/2$ one gets the contribution
$-Q_z(\tau_e^{-1}+\tau_h^{-1})$ to the time derivative of $N_b$ {, see Eq.~\eqref{dNb}.}
% Finally the commutator $\left[\mathcal H_x,\varrho(t)\right]$ gives the term $-Q_y\Omega$ describing electron-in-exciton spin precession.

The first two terms in the right hand side of the Eq.~\eqref{Sz} clearly describes electron spin precession and relaxation. Excitons spin flips and recombination also affect electron spin coherence. To describe this effect one should make a trivial decomposition
\[
 S_z=(3S_z-J_z)/6+(3S_z+J_z)/6,
\]
and note that the first term is zero for the dark excitons, while the second vanishes for the bright ones. Consequently these {terms} decay with the rates $1/\tau_b$ and $1/\tau_d$ respectively. Now the derivation of the rest of Eqs.~\eqref{dNbd}, \eqref{kinetic}, \eqref{kineticQ} is straightforward.


\begin{thebibliography}{10}
\bibitem{fnote} Additionally, the magnetic field $\Omega\sim\delta_0$ can produce an appreciable nonequilibrium electron spin polarization $S_x$ as well as the component $Q_y$.
\end{thebibliography}
\end{document}